\documentclass[11pt,twoside]{article}
\usepackage{asp2010}
\usepackage{graphicx}

\resetcounters

\begin{document}

\title{Modelling H$_2$ Infrared Emission of the Helix Nebula Cometary Knots}
\author{Isabel~Aleman$^{1,2}$, Albert~A.~Zijlstra$^1$, Mikako~Matsuura$^{3,4}$, 
       Ruth~Gruenwald$^2$, and Rafael~Kimura$^2$
\affil{$^1$Jodrell Bank Centre for Astrophysics, The Alan Turing Building, 
       School of Physics and Astronomy, The University of Manchester, 
       Oxford Rd, Manchester, M13 9PL, UK}
\affil{$^2$IAG-USP, Universidade de S\~ao Paulo, Cidade Universit\'aria, 
       Rua do Mat\~ao 1226, S\~ao Paulo, SP, 05508-090, Brazil}
\affil{$^3$UCL-Institute of Origins, Astrophysics Group, Department of 
       Physics and Astronomy, University College London, Gower Street,
       London, WC1E 6BT, UK}
\affil{$^4$UCL-Institute of Origins, Mullard Space Science Laboratory, 
      University College London, Holmbury St. Mary, Dorking, Surrey, 
      RH5 6NT, UK}
}

\begin{abstract}
In the present work, we use a photoionisation code to study the H$_2$ 
emission of the Helix nebula (NGC 7293) cometary knots, particularly that 
produced in the interface H$^+$/H$^0$ of the knot, where a significant 
fraction of the H$_2$ 1-0~S(1) emission seems to be produced. Our results 
show that the production of molecular hydrogen in such region may explain 
several characteristics of the observed emission, particularly the high 
excitation temperature of the H$_2$ infrared lines. 

\end{abstract}

\section{Introduction}

High-resolution images of the Helix nebula (NGC 7293) have shown that the 
H$_2$ emission arises from its large population of dense globules embedded 
in the ionised gas \citep[e.g.][]{Matsuura_etal_2009}, the so-called 
cometary knots (CKs). CKs are structures that resemble comets, particularly 
in images taken in H$\alpha$, [\mbox{N\,{\sc ii}}] $\lambda$6583, and 
H$_2$ 1-0 S(1) lines. The bright cusp points towards the central star and 
the tail in the opposite direction, which can indicate that the excitation 
is connected with the central star \citep{Odell_etal_2005, Odell_etal_2007}. 

The H$_2$ emission is intense in a thin layer on the surface of the CKs 
towards the central star. There is no evidence that this emission is 
produced by shocks \citep{Odell_etal_2005, Matsuura_etal_2007}. Models of 
photodissociation regions \citep[PDRs;][]{Cox_etal_1998,
Odell_etal_2007} are unable to reproduce the high excitation temperature  
of H$_2$ emission ($\sim$900-1800 K) estimated by \citet{Cox_etal_1998} 
and \citet{Matsuura_etal_2007}. Recently, \citet{Henney_etal_2007} showed 
that advection can cause the ionisation and dissociation front to merge, 
leading to enhanced heating of the molecular gas. 

In this work, we show that the partially ionised region can account for 
a significant part of the observed H$_2$ emission and naturally explain 
its high excitation temperature. Our models are briefly described in 
Section 2; a more detailed description will be published in a forthcoming 
paper \citep{Aleman_etal_2010}. Results are presented in Section 3.

\section{Models}

We use the photoionisation code {\it Aangaba\/} 
\citep{Gruenwald_Viegas_1992} to simulate the ionising spectrum, physical 
conditions, density of the species, and line emissivities around and 
inside the H$^+$/H$^0$ interface of the Helix nebula CKs. The H$_2$ 
micro-physics is included in the code, as described in 
\citet{Aleman_Gruenwald_2004,Aleman_Gruenwald_2010}. 

We assume that the central star radiates as a blackbody with $T_{\star}=$ 
120,000~K and $L_{\star}=$ 100 $L_{\sun}$ \citep{Henry_etal_1999, 
Odell_etal_2007}. We also assume that the density of the diffuse gas is 
uniform and equal to 50 cm$^{-3}$ \citep{Meixner_etal_2005}. 
The elemental abundances for the Helix were determined by 
\citet{Henry_etal_1999} for He, O, C, N, Ne, S, and Ar. For Mg, Si, 
Cl, and Fe, we adopt averages for PNe from \citet{Stasinska_Tylenda_1986}. 
We use amorphous carbon dust (with 0.1$\mu$m radius) for our calculations, but 
as discussed in \citet{Aleman_Gruenwald_2004} the choice of compound 
among the ones available in the code will not cause significant changes 
on the results presented below nor will affect our conclusions. 
The distance to the Helix is assumed to be 219 pc \citep{Harris_etal_2007}.
The CKs are simulated as an increase in the density profile of the Helix 
nebula model at a given distance from the central star. The emissivity 
along the radial direction (through the CK symmetry axis) is calculated 
by the photoionisation code. An IDL routine was developed to simulate a 
three-dimensional CK, allowing the calculation of line surface brightness 
by the integration of the emissivity in the line of sight inside a CK, 
which is assumed to be seen edge on. We construct a grid 
of CK models with different core densities, density profiles, dust-to-gas 
ratios, and distances from the central star. We assume that the density 
profile has a density increase from the diffuse gas to the CK core value 
within a given distance. We call this region the CK interface. We study four 
types of density increase with distance, but here we only present the 
results for the exponential increase \citep[more results will be included 
in][]{Aleman_etal_2010}. Calculation are stopped where the gas temperature, 
which generally decreases with the distance to 
the central star, reaches 100~K. In each model, the dust-to-gas ratio and 
the chemical composition of the CKs are assumed to be the same as in the 
diffuse gas.

\section{Results}

\subsection{Warm H$_2$ 1-0 S(1) emission}

The emissivity of the H$_2$ 1-0~S(1) line, as well as for other 
rovibrational lines, in the CKs is important in a warm region, where 
temperatures are between 300 and 7000~K. In the region considered 
in this work, the peak in the 1-0~S(1) emissivity occurs 
where the density is around 40\% of the core density.
The contribution of colder regions should be more important for pure 
rotational lines of the $v =$~0 level. This component of the H$_2$ 
emission may explain the excitation temperatures around 900-1800~K 
found by \citet{Cox_etal_1998} and \citet{Matsuura_etal_2007}. 

\begin{figure}
\centering
\includegraphics[width=120mm, clip=1]{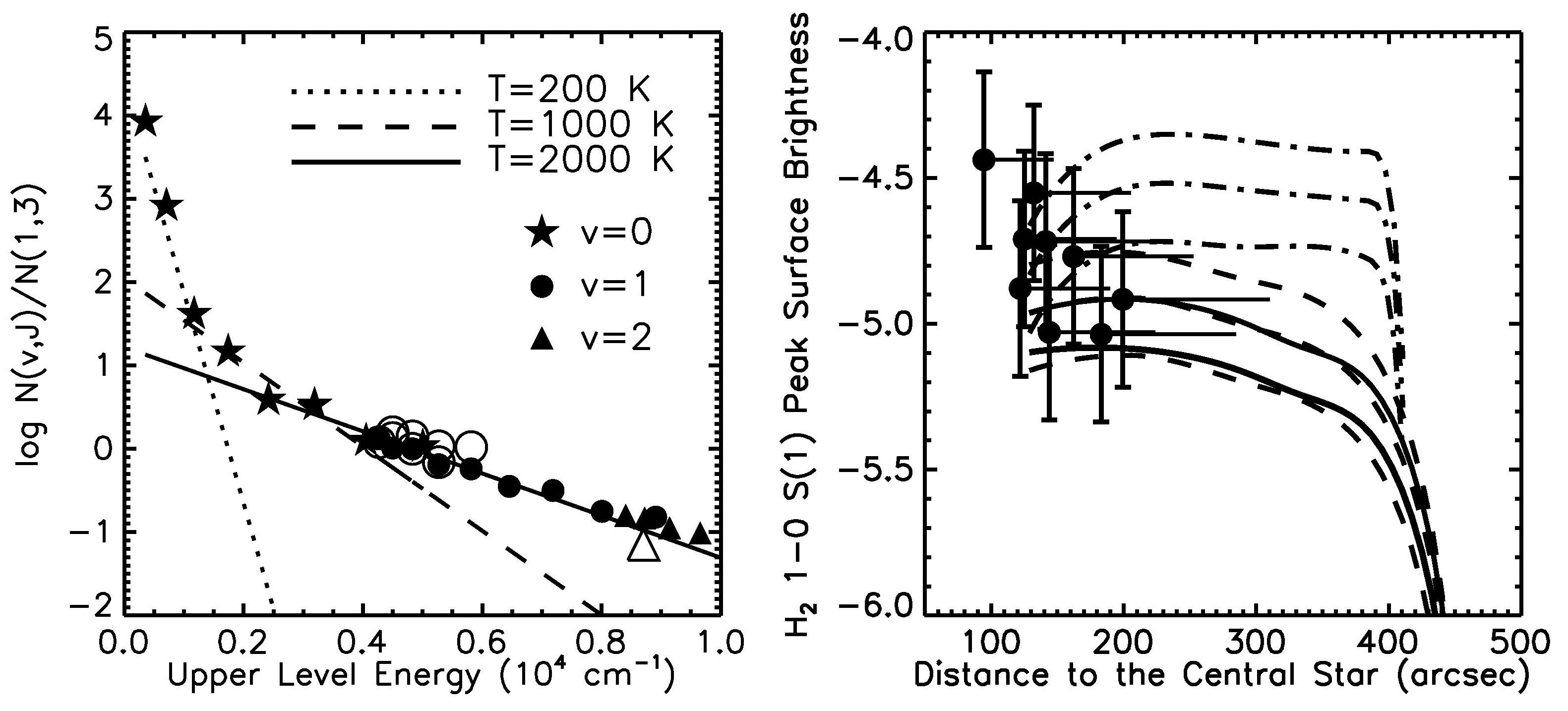}
\caption{Left Panel: H$_2$ excitation diagram. The effective column 
density was calculated where the 1-0~S(1) surface brightness is maximum. 
Open symbols represents observations and filled symbols 
models. Lines are Boltzmann distributions for the temperature indicated. 
Right panel: H$_2$ 1-0 S(1) surface brightness of a cometary knot as a 
function of its distance to the central star. Sets of solid, dashed, and 
dot-dashed curves represent models with interface thickness of 0.5, 0.2, and 
0.01\arcsec, respectively. Different curves within each set represent  
CK radius of 0.5, 1.0, and 2.0\arcsec, with the surface brightness 
increasing for larger CK radius. Dots represent measured values. 
The error in the distance is estimated assuming that the Helix 
symmetry axis is inclined 37$\,^{\circ}$ with respect to the line of sight}
\label{Fig1}
\end{figure}

The left panel in Fig. \ref{Fig1} shows the H$_2$ excitation diagram. 
Observational values were obtained by \citet{Matsuura_etal_2007}. 
Values calculated with an appropriate model are also 
included. The agreement between the excitation temperatures of the model 
and the observations is evident. Lines represent Boltzmann distributions 
for three different temperatures as indicated within the plot. 
The column densities 
obtained from the lines of the bands 1-0 and 2-1 are well 
represented by a excitation temperature of approximately 2000 K. A similar 
value was obtained by \citet{Matsuura_etal_2007}. The column densities 
obtained from lines 0-0 S(2) to S(7) are well represented by Boltzmann 
distribution at a temperature of 1000K, which is close to the 
excitation temperature of 900K obtained by \citet{Cox_etal_1998} from ISO 
observations of the Helix.

\subsection{H$_2$ 1-0 S(1) surface brightness} 

The right panel of Fig. \ref{Fig1} shows the H$_2$ 1-0 S(1) line surface 
brightness as a function of the CK distance to the central star. Dots 
represent measurements for some representative CKs of the Helix 
nebula. We identified 10 isolated CKs commonly detected in H$\alpha$ 
and H$_2$ images. 
We measured 2.12\,$\mu$m H$_2$ intensities from the images obtained by 
\citet{Matsuura_etal_2009}. To calibrate the data for point sources on the 
local scale, we use five stars within the observed field to measure the 
zero-pint. We assume that the 2MASS $K'$-magnitude of these stars are the 
same as the magnitudes in H$_2$-filter. We apply the 25-pixel radius of 
aperture photometry and take the 35--50 pixel ring as background 
measurements. The pixel scale is 0.117\arcsec.

Curves in the right panel of Fig. \ref{Fig1} show the H$_2$ 1-0 S(1) 
surface brightness as a function of the distance from the central star 
for CKs with different interface thickness and radius. The surface 
brightness was calculated was averaged over the same aperture as the 
measurements to allow direct comparison. The surface brightness 
decreases with the decrease of interface thickness, the increase of the 
CK radius, and the increase of the distance from the central star. 
The observed surface brightness also presents a decrease with distance to 
the central star trend. The interface can account for the whole or a 
significant part of the observed surface brightness. 

An important parameter to the ionization structure of the CKs is the 
distance from the central star, since the ionising spectrum may change 
significantly with the optical depth. CKs farther from the central star 
have smaller ionised zones. If the CK is beyond the Helix ionisation front, 
there is practically no ionised region and the intensity of 1-0~S(1) 
drops dramatically, since there is not enough radiation or temperature
to excite significantly the upper vibrational levels of the molecule.
Our results support that the central star's radiation field plays a 
major role in the CKs H$_2$ emission.

We also study the effect of 
the dust-to-gas ratio and $n_K$. The 1-0 S(1) peak brightness is slightly 
higher for models with higher dust-to-gas ratio (the increase caused by 
changing the dust-to-gas ratio from 10$^{-3}$ to 10$^{-2}$ is about 20\%) 
and higher $n_K$ (the difference between models with $n_K =$ 10$^{5}$ and 
10$^{6}$ cm$^{-3}$ is up to 40\%).

As pointed out by \citet{Burkert_Odell_1998}, the interface between the 
diffuse gas and the CK core may provide important clues about the 
mechanisms that shapes and sustain the CKs. Our models show that 
there are significant differences in the results depending on 
the assumed density profiles of this region. Images that could resolve 
the interface of the CK in great detail are then essential to improve 
or knowledge about the CKs and PNe.

\acknowledgements We acknowledge the financial support from CNPq 
PDE grant number 201950/2008-1 (Brazil), The University of 
Manchester, and STFC (UK). 

\bibliography{aleman}      

\end{document}